\documentclass{article}
\usepackage{amssymb}

\usepackage{graphicx}
\usepackage{amsmath}

\newtheorem{theorem}{Theorem}

\newtheorem{definition}[theorem]{Definition}

\newtheorem{remark}[theorem]{Remark}

\input{tcilatex}

\begin{document}

\title{Lightfront Formalism versus Holography\&Chiral Scanning\thanks{%
This work received support from the CNPq}}
\author{Bert Schroer \\
presently CBPF, Rua Dr. Xavier Sigaud, 22290-180 Rio de Janeiro, Brazil\\
email schroer@cbpf.br\\
permanent address: Institut f\"{u}r Theoretische Physik\\
FU-Berlin, Arnimallee 14, 14195 Berlin, Germany}
\date{August 2001\\
to appear in the proceedings of the 2nd International Symposium on ''Quantum
Theoriy and Symmetries'' Krakow, Poland \\
July 18-21, 2001}
\maketitle

\begin{abstract}
The limitations of the approach based on using fields restricted to the
lightfront (Lightfront Quantization or p$\rightarrow \infty $ Frame
Approach) which drive quantum fields towards canonical and ultimately free
fields are well known. Here we propose a new concept which does not suffer
from this limitation. It is based on a procedure which cannot be directly
formulated in terms of pointlike fields but requires ``holographic''
manipulations of the algebras generated by those fields. We illustrate the
new concepts in the setting of factorizable d=1+1 models and show that the
known fact of absence of ultraviolet problems in those models (in the
presence of higher than canonical dimensions) also passes to their
holographic images. In higher spacetime dimensions d\TEXTsymbol{>}1+1 the
holographic image lacks the transversal localizability; however this can be
remedied by doing holography on d-2 additional lightfronts which share one
lightray (Scanning by d-1 chiral conformal theories).
\end{abstract}

\section{A simple setting of the problem}

Lightfront quantum field theory and the closely related p$\rightarrow \infty 
$ frame method have a long history. The large number of articles on this
subject (which started at the beginning of the 70ies) may be separated into
two groups. On the one hand there are those papers whose aim is to show that
such concepts constitute a potentially useful enrichment of standard local
quantum physics \cite{Leut}\cite{DriesslerI}\cite{DriesslerII}\cite{Seiler},
but there are also innumerable attempts to use lightfront concepts as a
starting point of more free-floating ``effective'' approximation ideas whose
relations to causal and local quantum physics remained unclear. Here our
main interest is to extend the first mentioned results.

The main problem is to overcome the short distance limitations of the old
lightfront quantization and the p$\rightarrow \infty $ frame method (which
resulted in a severe restriction on the Kallen-Lehmann spectral function 
\cite{DriesslerI}) which unfortunately excludes all cases of genuine
renormalizable interactions. This restriction is identical to that which
canonical commutation relation (or the interpretation of the functional
measure in the spirit of a quantum mechanical Feynman-Kac euclidean
representation) requires. But whereas the (perturbative) breakdown of
canonicity causes no real harm (the canonical setting only serves as a
mental starter but plays no role in the Feynman formalism) since the more
general spacelike causality/locality structure is all one needs for dealing
with renormalization, the restriction for the lightfront approach is a more
serious matter. If the whole idea of lightfront-affiliated operators would
be limited to free fields as indicated in various rigorous investigations
about lightcone restrictions of Wightman fields\cite{DriesslerI}, the
subject would have only academic interest.

In a recent paper I have indicated how the field-coordinatization free
approach of AQFT can overcome this undesired restriction \cite{holo}. Its
intuitive basis is the radical different nature of the conformal structure
one encounters on the lightfront considered as a horizon of the wedge as
compared to the massless conformal scaling limit; in fact the correct way to
deal with lightfront localization is inexorably linked with the
understanding of wedge localized operator algebras.. Whereas the scaling
limit (which is closely linked to the renormalization group approach) admits
a natural formulation in terms of pointlike fields, the present chiral
conformal structure on a lightfront horizon of a wedge region associated
with interacting fields with nonintegrable Kallen-Lehmann spectral function
cannot be obtained simply as a restriction of such fields to the horizon.
One rather needs to perform a reprocessing of the field content into the net
of algebras and to apply a modular inclusion procedure to the wedge algebra.
In fact this process just supplies different spacetime indexed subalgebras;
as global algebras the chiral horizon algebra is identical with the wedge
algebra. Different from a restriction process, the notion of transversal
localization is lost (i.e. the obtained chiral theory is a kind of
kinematical coarse grained object) and has to be constructed by tilting the
wedge as described in section 4.

To physicists familiar with the Unruh-Hawking effect the special role of the
wedge algebra may not come totally unexpected, but the crucial role of
operator algebras versus pointlike fields is somewhat surprising in view of
the widespread opinion that their use is more a matter of mathematical
conciseness and conceptual clarity rather than a matter of principle.
Although concepts as Hawking temperature, horizons and holography originated
with black hole physics, they have their natural analogs in Minkowski space%
\footnote{%
Whereas in flat spacetime these local quantum physical concepts remain
hidden and play no important role in particle physics, curved spacetime in
special circumstances exposes them in a geometric quasiclassical veil.(in
which they were discovered). The claims that holography is a characteristic
manifestation of quantum gravity should be treated with cautious scepticism.
In fact the present paper casts doubts on suggestions that such concepts
require new physical principles which go beyond those of local quantum
physics; as the situation looks now it appears more and more like a lack of
agreement between older conjectures versus more recent theorems.}. Whereas
for thermal aspects this had been known for some time, the presence of the
other concepts is the main issue of this article.

Although our conceptual setting and the mathematical formalism goes beyond
what the protagonist of \ ``holographic projections'' onto the horizon had
in mind \cite{Hooft} and how this became subsequently related with
lightfront quantization ideas \cite{Susskind}, the intuitive physical
content of our algebraic approach matches these ideas perfectly. For this
reason (and also for the fact that we cannot think of any other way in which
this idea could be furnished with a rigorous mathematical and conceptually
tight local quantum physical meaning) the excellent name is maintained; in
fact we think of our conceptual additions as the liberation of these ideas
from their old short distance restrictions which did not leave any room for
interactions.

The name holography is often used in a wider sense for an isomorphism of a
local (massive) QFT with data which are associated with lower spacetime
dimensions (In case of equal spacetime dimensions of the source and target
theory the name transplantation has been used \cite{Buch}). Not every aspect
of such an isomorphism requires the full mathematical power of the field
coordinatization free formulation of AQFT. For example the direction 
\begin{equation*}
AdS\rightarrow CQFT
\end{equation*}
in the Maldacena conjecture \cite{Maldacena} can be seen by taking the limit
of the spacial infinite remote Anti deSitter boundary in correlation
functions of fields and is therefore susceptible to standard formulations of
QFT. Only the existence of the inverse and the understanding of the detailed
properties which one field theory inherits from the other requires a more
ambitious mathematical theorem in the AQFT setting\footnote{%
One rigorous result is that the Maldacena conjecture cannot hold between two
Lagrangian (pointlike fields) quantum field theories \cite{Rehren}\cite
{Schroer}\cite{Smolin}.} \cite{Rehren}. In the present case of lightray
holography the algebraic setting is crucial for both directions of the
isomorphism. Holography in this sense (where localizations becomes scrambled
up) is not useful for physical concepts which rely on spacetime localization
between operators as e.g. time dependent scattering theory; rather their
beneficial effect is expected to show up in the understanding of
coarse-grain properties of a localized algebra as e.g. its relative
localization entropy between different states where only the overall
relative size of degrees of freedom matter, but not the sunlocalizations
within that algebra. These matters will be discussed in a forthcoming paper.

The content is organized as follows. In the next section the properties of
lightray restrictions in d=1+1 are recalled in the present setting. Section
3 illustrates the new construction based on modular wedge localization
within the rich family of factorizing d=1+1 models. The additional problems
posed by higher dimensional QFT including proposals for their solution are
then addressed in section 4. The last section attempts to make connections
to other fundamental problems of QFT for whose solution the new concepts are
also expected to be important. In a mathematical appendix we collect some
known mathematical results for the convenience of the reader.

\section{Elementary review of d=1+1 lightray restriction}

Let us consider the simplest case namely the lightray holography of a
two-dimensional massive selfconjugate QFT. We first remind ourselves of the
situation for a free massive scalar field 
\begin{eqnarray}
A(x) &=&\frac{1}{\sqrt{2\pi }}\int \left( e^{-ipx}a(\theta )+e^{ipx}a^{\ast
}(\theta )\right) d\theta \\
p &=&m(ch\theta ,sh\theta )  \notag
\end{eqnarray}
where for convenience we used the momentum space rapidity description. In
order to approach the light ray $x_{-}=t-x=0$ in such a way that $x_{+}=t+x$
remains finite we approach the $x_{+}>0$ horizon of the right wedge $%
t^{2}-x^{2}<0,$ $x>0$ by taking the $r\rightarrow 0,\,\chi =\hat{\chi}%
-lnr\rightarrow \hat{\chi}$ +$\infty $ in the x-space rapidity
parametrization 
\begin{eqnarray}
&&x=r(sh\chi ,ch\chi ),\,\,x\rightarrow (x_{-}=0,x_{+}\geq 0,\text{ }finite)
\label{lim} \\
&&A(x_{+},x_{-}\rightarrow 0)\equiv A_{+}(x_{+})=\frac{1}{\sqrt{2\pi }}\int
\left( e^{-ip_{-}x_{+}}a(\theta )+e^{ip_{-}x_{+}}a^{\ast }(\theta )\right)
d\theta  \notag \\
&=&\frac{1}{\sqrt{2\pi }}\int \left(
e^{-ip_{-}x_{+}}a(p)+e^{ip_{-}x_{+}}a^{\ast }(p)\right) \frac{dp}{\left|
p\right| }  \notag
\end{eqnarray}
where the last formula serves to make manifest that the limiting $A(x_{+})$
field is a chiral conformal (gapless $P_{-}$ spectrum) field; the mass in
the exponent $p_{-}x_{+}=mre^{\theta }e^{-\chi }$ is just a parameter which
keeps track of the ``engineering dimension'' (the physical mass is the gap
in the $P_{-}\cdot P_{+}$ spectrum). Since this limit only effects the
numerical factors and not the Fock space operators $a^{^{\#}}(\theta ),$ we
expect that there will be no problem with the horizontal restriction i.e.
that the formal method (the last line in \ref{lim}) agrees with the rigorous
result. Up to a fine point which is related to the bad infrared behavior of
a scalar $dimA=0,$ field this is indeed the case. Using the limiting $\chi $%
-parametrization we see that for the smeared field with $supp\tilde{f}\in W$
one has the identity 
\begin{eqnarray}
&&\int A(x_{+},x_{-})\tilde{f}(x)d^{2}x=\int_{C}a(\theta )f(\theta )=\int
A_{+}(x_{+})\tilde{g}(x_{+})dx_{+} \\
&&\tilde{f}(x)=\int_{C}e^{ip(\theta )x}f(\theta )d\theta ,\,\,\tilde{g}%
(x_{+})=\int_{C}f(\theta )e^{ip_{-}(\theta )x_{+}}d\theta  \notag
\end{eqnarray}
These formulas warrant some explanation. The onshell character of free
fields restricts the Fourier-transformed test function to their mass shell
values $f(p)|_{p^{2}=m^{2}}=f(\theta )$ and the wedge support property is
equivalent to the strip analyticity. The integration path $C$ consists of
the upper and lower rim of the strip and corresponds to the
negative/positive frequency part of the Fourier transform. By introducing
the test function $\tilde{g}(x_{+})$ which is supported on the halfline $%
x_{+}\geq 0$ it becomes manifest that the smeared field on the horizon
rewritten in terms of the original Fourier transforms must vanish at $p=0$
since 
\begin{eqnarray*}
f(p)|_{p^{2}=m^{2}}\frac{dp}{\sqrt{p^{2}+m^{2}}} &=&f(\theta )d\theta \equiv
g(\theta )d\theta =g(p)\frac{dp}{\left| p\right| } \\
&\curvearrowright &g(0)=0
\end{eqnarray*}
This infrared restriction is typical for $dimA=0$ fields and would not occur
for free fields with a nontrivial L-spin. The equality of the f-smeared $%
A(x) $ fields with the g-smeared $A_{+}(x_{+})$ leads to (after
clarification of some domain problems of these unbounded operators) the
equality of the affiliated (weakly closed) operator algebras 
\begin{equation*}
\mathcal{A}(W)=\mathcal{A}(R_{+})=\mathcal{A}(R_{-})
\end{equation*}
Here the last equality expresses the fact we could have taken the lower
horizon with the same result. This equality is the quantum version of the
classical propagation property of characteristic data on the upper or lower
lightfront of a wedge. With the exception of d=1+1 m=0 the amplitudes inside
the causal shadow $W$ of $R_{+}$ are uniquely determined by the lightfront
data. Note that a finite interval on $R_{+}$ does not cast a 2-dimensional
causal shadow; this is only the case for the full characteristic data.
Related to this is fact that the opposite lightray translation 
\begin{eqnarray*}
AdU_{-}(a)\mathcal{A}(R_{+}) &\subset &A(R_{+}) \\
U_{-}(a) &=&e^{-iP_{-}a}
\end{eqnarray*}
is a totally fuzzy endomorphism of the $A(R_{+})\,$net whereas in the
setting of the $\mathcal{A}(W)$ net spacetime indexing it is a geometric map.

It is very important to notice that even in the free case the horizontal
limit is different from the scale invariant massless limit. The latter
cannot be performed in the same Hilbert space since the $m\rightarrow 0$
limit needs a compensating $\ln m$ term in the momentum space rapidity $%
\theta $ of the operators $a^{\#}(\theta )$ whereas in the horizon limit it
was only effecting the c-number factors. There is no problem of taking this
massless limit in correlation functions if one uses smearing functions whose
integral vanishes (or $f(p=0)=0$). The limiting correlation functions define
via the GNS construction a new Hilbert space which contains two chiral
copies of the conformal dimA=0 field corresponding to the right/left movers.
In the case of interacting theories the appropriately defined horizontal
holographic projection is different from the scaling limit by much more than
just multiplicities as will be seen below.

For interacting fields a necessary condition for the above lightray
restriction to work is the finiteness of the wave function renormalization
constant which in terms of two-point function of the correctly normalized ($%
\left\langle p\left| A(0)\right| \Omega \right\rangle =\frac{1}{\sqrt{2\pi }}%
)$ interpolating field is the convergence of the following integral over the
Kallen-Lehmann spectral function 
\begin{equation}
\int \rho (\kappa ^{2})d\kappa ^{2}<0  \label{finite}
\end{equation}
This condition is violated for all genuinely renormalizable (i.e. not
superrenormalizable) theories. In this case the holographic reprocessing has
to be done on the level of algebras according to the formal scheme 
\begin{equation*}
A(x)\rightarrow \mathcal{A}(W)\,net\overset{holography}{\longrightarrow }%
\mathcal{A}(R_{+})\,net\rightarrow A_{+}(x_{+})
\end{equation*}
Here $A(x)$ stands symbolically for a complete set of local fields which
fulfill d=1+1 locality and $A_{+}(x_{+})$ stand for possible $R_{+}$-local
chiral fields.

\section{Holography for factorizable models}

The holography in the presence of renormalizable interactions which violate
the finiteness condition (\ref{finite}) can be nicely illustrated for d=1+1
factorizable models. The most appropriate construction of these models
starts from free fields with nonlocally modified commutations relations for
the momentum space creation and annihilation operators \cite{JMP} 
\begin{eqnarray*}
A(x) &=&\frac{1}{\sqrt{2\pi }}\int \left( e^{-ipx}Z(\theta )+e^{ipx}Z^{\ast
}(\theta )\right) d\theta ,\,\,p=m(ch\theta ,sh\theta ) \\
A(\tilde{f}) &=&\frac{1}{\sqrt{2\pi }}\int_{C}Z(\theta )f(\theta )d\theta
\end{eqnarray*}
where the Z's are defined on the incoming n-particle vectors by the
following formula for the action of $Z^{\ast }(\theta )$ for the
rapidity-ordering $\theta _{i}>\theta >\theta _{i+1},\,\,\theta _{1}>\theta
_{2}>...>\theta _{n}$ 
\begin{align}
& Z^{\ast }(\theta )a^{\ast }(\theta _{1})...a^{\ast }(\theta
_{i})...a^{\ast }(\theta _{n})\Omega = \\
& S(\theta -\theta _{1})...S(\theta -\theta _{i})a^{\ast }(\theta
_{1})...a^{\ast }(\theta _{i})a^{\ast }(\theta )...a^{\ast }(\theta
_{n})\Omega  \notag \\
& +contr.\,from\text{ }bound\,states  \notag
\end{align}
In the absence of bound states this amounts to the commutation relations

\begin{align}
Z^{\ast }(\theta )Z^{\ast }(\theta ^{\prime })& =S(\theta -\theta ^{\prime
})Z^{\ast }(\theta ^{\prime })Z^{\ast }(\theta ),\,\theta <\theta ^{\prime }
\label{ab} \\
Z(\theta )Z^{\ast }(\theta ^{\prime })& =S(\theta ^{\prime }-\theta )Z^{\ast
}(\theta ^{\prime })Z(\theta )+\delta (\theta -\theta ^{\prime })  \notag
\end{align}

A smeared field $A(\hat{f})$ applied to the vacuum creates a one-particle
vector 
\begin{equation*}
A(\hat{f})\left| 0\right\rangle =\int f(\theta -i\pi )\left| p(\theta
)\right\rangle d\theta
\end{equation*}
Localized operators which applied to the vacuum create one-particle vectors
without admixture of multiparticle vacuum polarization clouds are called
polarization-free generators (PFG's). Their structure is restricted by the
following theorems \cite{JMP}

\begin{theorem}
PFG's which are localized in regions whose causal completion is genuinely
smaller than a wedge (e.g. spacelike cones, double cones) lead to
interaction-free theories. On the other hand wedge-localized PFG's always
exist even in the presence of interactions. They have Fourier transforms
(tempered distributions) only in the case of d=1+1 with purely elastic
scattering.
\end{theorem}

According to this theorem we should ask the question whether the above PFG's
are wedge-localized. The affirmative answer is contained in the following
theorem

\begin{theorem}
The PFG's with the above algebraic structure for the Z's are wedge-localized
if and only if the structure coefficients S($\theta $) are meromorphic
functions which fulfill crossing symmetry in the physical $\theta $-strip
i.e. the requirement of wedge localization converts the Z-algebra into a
Zamolodchikov-Faddeev algebra.
\end{theorem}

In this case the $A(\hat{f})$ are generators affiliated to the wedge algebra 
\begin{equation*}
\mathcal{A}(W)=alg\left\{ A(f)|suppf\subset W\right\}
\end{equation*}
The most general operator $A$ in $\mathcal{A}(W)$ is a LSZ-type power series
in the Wick-ordered Z's

\begin{equation}
A=\sum \frac{1}{n!}\int_{C}...\int_{C}a_{n}(\theta _{1},...\theta
_{n}):Z(\theta _{1})...Z(\theta _{n})d\theta _{1}...d\theta _{n}:
\label{series}
\end{equation}
with strip-analytic coefficient functions $a_{n}$ which are related to the
matrix elements of $A$ between incoming ket and outgoing bra multiparticle
state vectors. The integration path $C$ consists of the real axis
(associated with annihilation operators and the line $Im\theta =-i\pi .$
Writing such power series without paying attention to domains of operators
means that we are we are only dealing with bilinear forms whose operator
status is still to be settled.. The bilinear forms which have their
localization in double cones are characterized by their relative commutance
(this formulation has to be changed for Fermions or more general objects)
with shifted generators $A^{(a)}(f)\equiv U(a)A(f)U^{\ast }(a)$%
\begin{eqnarray*}
\left[ A,A^{(a)}(f)\right] &=&0,\,\forall f\,\,suppf\subset W\, \\
A &\subset &\mathcal{A}_{bil.}(C_{a})
\end{eqnarray*}
where the subscript indicates that we are dealing with spaces of bilinear
forms (formfactors of would-be operators localized in $C_{a}$) and not yet
with unbounded operators and their affiliated von Neumann algebras. This
relative commutant relation \cite{JMP} on the level of bilinear forms is
nothing but the famous ''kinematical pole relations'' which relate the even $%
a_{n}$ to the residuum of a certain pole in the $a_{n+2}$ meromorphic
functions. The structure of these equations is the same as that for the
formfactors of pointlike fields; but whereas the latter lead (after
splitting off common factors \cite{Karowski} which are independent of the
chosen field in the same superselection sector) to polynomial expressions
with hard to control asymptotic behavior, the $a_{n}$ of the double cone
localized bilinear forms are solutions which have better asymptotic behavior
which according to the Payley-Wiener-Schwartz theorem. We will not discuss
here the problem of how this improvement can be used in order to convert the
bilinear forms into genuine operators.

Suppose now that we consider a modular inclusion (see appendix) of wedges
which in the above formalism just means that we take $a$ as a lightlike
vector $a\sim (1,1).$ In that case the relative commutant consists of
bilinear forms which are interpreted to be localized in the interval $(0,1)$
on the lightray which according to general theorems are associated to a
chiral conformal field theory. It is easily checked that the previous space
of double cone localized space is reobtained (as expected) by applying a
suitable opposite lightray translation to the interval localized space of
bilinear forms. The total spaces of wedge-localized and lightray localized
bilinear forms (\ref{series}) which are the (weak closures of the) unions of
the local spaces are the same (as expected from the classical picture that
the causal shadow of the characteristic data on the halfline is the wedge
i.e. $\mathcal{A}(R_{+})=\mathcal{A}(W)$ whereas finite intervals cast no
shadow).

According to the construction the net on the horizon is bosonic and since it
is also chiral the spectrum of scale dimensions must be integer-valued. The
massless limit of the pointlike field generators of the wedge theory on the
other hand can have anomalous short distance behavior. This is yet another
manifestation of the useful kinematical nature of the lightray algebra (the
``scrambled up'' wedge algebra). Important dynamical data have been
transferred to the action of the opposite lightray translation which plays
the role of a kind of hamiltonian and whose action destroys the conformal
invariance and recreates the complications of the original massive theory.
The fascinating aspects of the free field behavior of pointlike generators
of the chiral $\mathcal{A}(R)$ net is the extreme nonlocality they must have
with respect to the generators of the $\mathcal{A}(W)$ net together with the
naturalness of their construction. Despite their simplicity their
creation/annihilation operators must be infinite power series in the $Z$%
-operators. Factorizing models promise to play an important role in the
better understanding of this horizont-localized chiral conformal field
theory which is quite different from the conformal scaling limit theory
(e.g. it has a richer supply of symmetries and maps). Automorphisms which
act locally on the original set may become ``fuzzy'' on the horizon net
(example: the opposite lightray translation) and vice versa (example the
circular rotation on the conformal horizon).

\section{Problems met in higher dimensional cases}

When there are transversal spatial dimensions the restriction of fields to a
lightfront continues to show the same problems in the presence of
interactions. For a proof that (under very mild assumptions about operator
domains) the fields must have the free canonical structure we refer to the
second paper of Driessler. In the first paper the author also shows that the
vacuum factorizes transversally\footnote{%
This observation is closely related to the impossibility to associate an
operator algebra on a lightray in the presence of transverse directions.
\par
{}} (a behavior otherwise met for spatially disjoint localized algebras in
theories without vacuum polarization like second quantized QM) although it
remains highly entangled with respect to longitudinal (along the light ray)
disjointness where one needs the so-called split property in order to meet
conditions which are similar to quantum mechanical (type I) algebras.

Before we turn to a field-free i.e. algebraic construction of a holographic
projection we briefly present the lightfront formalism for free fields. The
rapidity parametrization of a scalar free field with $x=r(sh\chi ,ch\chi
,x_{\perp }),\,\,p=(m_{eff}ch\theta ,m_{eff}sh\theta ,p_{\perp }),$ $%
m_{eff}(p_{\perp }^{2})=\sqrt{m^{2}+p_{\perp }^{2}}$ leads to the following
(upper) horizontal projection formula 
\begin{eqnarray}
A(x_{+},x_{\perp }) &=&\frac{1}{\left( 2\pi \right) ^{\frac{d}{2}}}\int
(e^{-ip_{-}x_{+}+ip_{\perp }x_{\perp }}a(\theta ,p_{\perp
})+e^{ip_{-}x_{+}-ip_{\perp }x_{\perp }}a^{\ast }(\theta ,p_{\perp }))d\theta
\notag \\
p_{-}x_{+} &=&m_{eff}(p_{\perp }^{2})e^{\theta }x_{+}
\end{eqnarray}
Rewritten formally in terms of the momentum space measure 
\begin{equation}
A(x_{+},x_{\perp })=\frac{1}{\left( 2\pi \right) ^{\frac{d}{2}}}\int
(e^{-ip_{-}x_{+}+ip_{\perp }x_{\perp }}a(p_{-},p_{\perp
})+e^{ip_{-}x_{+}-ip_{\perp }x_{\perp }}a^{\ast }(p_{-},p_{\perp }))\frac{%
dp_{-}}{\left| p_{-}\right| }
\end{equation}
we encounter the typical infrared divergence which requires to use again the
restricted test functions $f(x_{+},x_{\perp })$ (again only for s=0) with $%
\int f(x_{+},x_{\perp })dx_{+}=0$. The above rapidity representation reveals
that the lightfront field almost looks like a continuous superposition of
chiral fields at different scales set by the continuous values of the
magnitude of the transverse momenta $\left| p_{\perp }\right| .$

As in the d=1+1 case in section 3, the algebraic holography starts from the
modular inclusion of the standard wedge shifted along the $x_{+}$ lightray
into itself (the reason we work with wedge algebras and not with the full
algebra is that the modular prerequesites of absence of annihilation
operators for the vacuum are violated for the latter). Unlike the lightfront
restriction of free fields which together with the action of the opposite
lightray translation allows to reconstruct the original d=1+1 theory, the
modular inclusion in the present case yields a result which has no
transverse localization, although the transverse translation $e^{iP_{\perp
}x_{\perp }}$ acts on $A(R_{+}).$ Note that our notation is not meant to
literally indicate a localization on the $x_{+}$ lightray but only indicates
that the chiral holographic projection has no transversal localization
concept (otherwise we could not describe it in terms of a chiral theory).
This is the prize we have to pay for obtaining such a simple holographic
image from a complicated higher dimensional QFT. Knowing the action of the
translation into the opposite lightray direction does not help to resolve
the lack of transversal localization. Whereas this may be enough in problems
of degree of freedom counting (e.g. entropy discussions) for the
reconstruction of the original net i.e. for the formulation of a holographic
isomorphism one must get a control of transversal localization. This can be
achieved by Lorentz tilting the standard wedge around its $x_{+}$ lightray.
This is done by using the ``translational'' part of the Wigner little group
of the light ray. In d dimensions there are d-2 such translations inside the
homogenous Lorentz group and they act like a transversal Galilei
transformation. Together with the longitudinal symmetries the geometric
symmetry group of a lightfront is seven-dimensional \cite{Leut}: the
longitudinal dilation (alias L-boost) and translation and the two
transversal translations as well as Galilei transformations and a
transversal rotation (the wedge allows a natural association with an
8-parametric subgroup of the Poincar\'{e} group \cite{S-W}).

In order to keep things geometrically simple let us choose in the following
d=1+2. Then there is just one one-parametric group of tilting (Galileian)
transformations such that the pair of the original wedge together with the
tilted wedge forms a modular intersection in the sense of the appendix. This
may be easily rephrased in terms of an additional localization structure on
the upper lightfront horizon. The original transversal stripes which
correspond to the longitudinal intervals are transformed into sloped stripes
which intersect the original ones in parallelograms and form the
localization regions of an algebraic net on the horizon. Again this is a
fuzzy transformation of the chiral theory. Clearly the parallel translates
of the tilted lightfront intersect the original lightfront in lines which
are transversally shifted parallels of the original lightray. This can be
used to generate a transversal localization on the original lightfront which
generates a net structure on this lightfront. Together with the action of
the opposite lightray translation this is sufficient to recuperate the
original d=1+2 net. Instead one may also talk about reconstructing the
original theory by scanning with two chiral theories, the second one
resulting from the first by the tilting automorphism. The principle of
generalization to d dimensions should be clear from our geometric
interpretation.

\section{Hopes based on modular localization}

In these notes we have shown that (in the presence of interactions) modular
localization and modular inclusions are essential tools in the formulation
of ideas around ``holography'' as a significant extension of lightcone
quantization. A closely related problem is that of modular symmetries beyond
the geometric symmetries of Poincar\'{e}- or conformal invariance (which are
also known to be of modular origin, in the case of the infinite chiral
diffeomorphisms group this was shown in \cite{Fuzzy}). Fuzzy modular groups $%
\sigma _{t}$ exist for each standard Reeh-Schlieder pair ($\mathcal{A}(%
\mathcal{O}),\Phi ),$ they only depend on the state $\phi $ and not on its
implementing vector $\Phi $. They are consistent with causality because they
keep the causal complement of $\mathcal{O}$ apart from $\mathcal{O}$ in
analogy say to the wedge affiliated L-boost which does not mix $\mathcal{A}%
(W)$ observables with those of $\mathcal{A}(W^{\prime })=\mathcal{A}%
(W)^{\prime }$ and as globally defined automorphisms they maintain
commutance if applied to originally commuting operators even if the original
localization regions suffer a fuzzy t-dependent ``diffusion''. They do not
exist in the corresponding classical theory (assuming that the
correspondence principle creates such an object) and their concrete form
depends much more on quantum dynamical aspects. The intrinsic understanding
of interactions is related to the absence of better than wedge localized
PFG's and therefore it seems to be a reasonable conjecture that the action
of the modular groups on vectors created by the application of such PFG to
the vacuum is related to the shape of their vacuum-polarization clouds. An
intrinsic understanding of these interaction-caused clouds is of course an
old but never fulfilled dreams of nonperturbative local quantum physics.

Another such dream (which was mentioned in the introduction) is the
understanding of ``entropy of localization'' \cite{Narn}\cite{JMP}; this is
natural problem since ``localization temperature'' (the Unruh-Hawking
effect) has already been successfully explained in terms of modular concepts 
\cite{Sewell}. In particular it would be of great interest to know whether
localization entropy preempts the Bekenstein black hole behavior and is
proportional to the area of the spatial boundary of the bounding causal
horizon and if the analogs to the thermodynamical \ laws are in some sense
quantum renormalized.

The formalism for factorizable models which we used for d=1+1 illustrations
of holography in section 2 is of course interesting in its own right, since
it offers a constructive approach without facing any ultraviolet problem
just as it was expected way back in a pure S-matrix approach. Its pointlike
formulation \cite{Smir} has witnessed a quite impressive growth over many
years and the present operator algebraic method adds to it a spacetime
interpretation of the Smirnov axioms (in particular of the
Zamolodchikov-Faddeev nonlocal creation and annihilation operators) and
holds the promise of a construction of the operators behind the bilinear
forms (formfactors). Operator constructions are easier for extended objects
than for pointlike fields.

The insensitivity of modular-based concepts on ultraviolet behavior of
pointlike fields calls for the reinvestigation of the perhaps most important
question of the post Feynman era: does the standard renormalizabilty
criterion really mark the short distance frontier of the underlying
principles of local quantum physics or does it only limit the range of
applicability of trying to understand interactions in terms of singular
pointlike field coordinatizations? Whereas in the Lagrangian renormalization
approach based on power counting massive free fields cannot have dimensions
beyond one in d=1+3, the bootstrap-formfactor approach, wherever it works,
does not know such bounds.

Unfortunately this onshell approach remains presently limited to factorizing
models in d=1+1 where it constructs a tempered (polynomially bounded in
momentum space) field theory \cite{BBS} for each admissable (crossing
symmetric, unitary) factorizing S-matrix; no a priory restriction on short
distance behavior is necessary. This situation was what the S-matrix
protagonists dreamed of in the 60ies \cite{Weiss}. Although they extracted
most of their S-matrix properties (in particular the important crossing
property) from the causality and spectral properties of local quantum
physics they had the unfortunate idea that in order to be successful in
their pursuit they had to liquidate QFT. One could of course adapt the
pessimistic attitude that these factorizing models are too special since
they all share the property that whenever they contain a variable coupling
parameter (example: massive Thirring model) there is no genuine coupling
constant renormalization and the Beta function vanishes.

There are however also encouraging indications from the perturbative aspects
of higher spin theories whose couplings violate the orthodox power counting
criterion. Consider for example interactions of massive d=1+3 vectormesons.
Since the free field dimensions of the vector description is two instead of
one (other associations of physical free fields with s=1 cannot lower the
dimension), any Lorentz-invariant interaction involves has at least the
powercounting dimension 5 and hence to the orthodox approach it is
nonrenormalizable i.e. the number of counterterm structures increase in each
order of perturbation theory which leads also to a nontempered behavior of
correlation functions. On the other hand there is the following trick \cite
{Duetsch} (only available for massive vectorfields) which leads to a
renormalizable solution \ i.e. one for which the operator dimension 2 of the
vectormeson field (apart from the expected logarithmic corrections) is
maintained after renormalization. The trick consists of representing the
Wigner one particle space as a cohomology space in the spirit of a one
particle version of the so-called BRST formalism and constitutes a
generalization of the approach taken in \cite{Scharf}. The idea behind this
trick is the same as in any application of BRST formalism: the combination
of lowering the propagator dimensions in the cohomological extension
together with the idea of stability of cohomological representations allows
a safe return to physics after having renormalized the auxiliary correlation
function of the extended correlation functions in the standard way
(perturbative renormalization is a linear process which does not require
positivity). The particular ``one-particle version'' of the BRST
representation has two additional advantages. On the one hand it keeps the
BRST ``ghosts'' away from self-interactions (their contribution to the
action remains bilinear) and maintains the LSZ asymptotic structure
throughout. On the other hand it requires the presence of additional
interacting physical degrees of freedom which (unlike in the standard
approach based on the mass generation by Higgs condensates) was not part of
the input. The simplest (and probably only) realization of these additional
degrees of freedom is via a standard scalar field i.e. one with vanishing
vacuum expectations (i.e. not ``Higgs'' if this terminology refers to the
condensate). There is of course no change of physical content as compared to
the standard Higgs mechanism approach, but the perspective and physical
interpretation is somewhat different. In particular one returns to the
original physical question for renormalizable vectormesons in the pre-gauge
spirit of Sakurai and Lewellyn-Smith. In this way the suspicion that there
was only one renormalizable coupling involving massive vectormesons is
confirmed and a gauge principle which selects between different possibilities%
\footnote{%
There are however several couplings involving (semi)classical vector fields.
In that case the gauge principle selects the Maxwellian interactions. To the
extend that the correspondence principle can be invoked, the classical gauge
principle follows from renormalizability of the corresponding local quantum
physics.} is not needed. From a particle physics point of view it makes more
sense to emphasize the uniqueness of \ renormalizable vectormeson couplings
than the differential geometric esthetical appeal of the gauge principle in
the Higgs mechanism. One hopes that the present modular concepts may have
will tell us eventually something about the true borders carved out by the
underlying physical principles and perhaps show that the widening of
renormalizability does not stop at the interaction of spin s=1 particles.

Another more mundane goal where one expects the new concepts to produce
explicit answers is the local quantum physics of d=1+2 anyons and of the
d=1+3 infinite spin zero mass Wigner representations. In both cases the
multiparticle structure does not follow the usual tensor product structure
and there exist no PFG's which have a better than wedge-like localization 
\cite{Mund}. In fact the maintenance of the spin-statistics connection in
the presence of braid group statistics requires even in the ``free'' case
(conservation of real particles, even vanishing of the cross section) the
presence of vacuum polarization clouds so that the nonrelativistic limit
remains a QFT i.e. cannot be a QM. This connection is in particular not
maintained in the QM of the Aharonov-Bohm like topological constructions.

We hope that we were able to convince the reader that there are plenty of
deep and physically relevant local quantum physical problems in particle
physics which fall into the range of the new method.

\textit{Acknowledgements:} Some helpful informations by Detlev Buchholz
concerning the older literature are greatfully acknowledges. The persuit of
content of old papers also brought about an enjoyable electronic
correspondence with Heinrich Leutwyler, a colleague and friend from the old
``current algebra and lightcone days''. He confirmed that the simplicity of
lightfront algebras was the prime motive for their introduction, a
motivation which the modern holography inherited.

\section{Appendix: mathematical aspects of modular theory}

For the benefit of the uninitiated reader we briefly collect and comment on
some formulas from modular theory \cite{Bo}.

Let ($\mathcal{A},\Omega $) be a standard pair, i.e. an operator algebra in
a Hilbert space $H$ with a vector $\Omega $ on which the algebra acts
cyclically ($\overline{\mathcal{A}\Omega }=H$) and on which there exist no
annihilators ($A\Omega =0,\,A\in \mathcal{A\curvearrowright \,}A=0,$ ``$%
\Omega $ is separating'')

\begin{definition}
Tomita's involution: SA$\Omega =A^{\ast }\Omega ,\curvearrowright S$ is
closed, antilinear and involutive i.e. S$^{2}\subset \mathbf{1}$ (involutive
on the domain). The polar decomposition $S=J\Delta ^{\frac{1}{2}}$ leads to
an antiunitary $J$ and a positive $\Delta $ which in turn gives rise to a
one parametric unitary group$\Delta ^{it}$, the (Tomita) modular group
\end{definition}

\begin{theorem}
(Tomita, Takesaki) The Ad-action of $\Delta ^{it}$ defines an automorphism $%
\sigma _{t}$ of the operator algebra $\mathcal{A}$ and the Ad -action of $J$
maps $\mathcal{A}$ into its commutant algebra $\mathcal{A}^{\prime }$%
\begin{eqnarray}
\sigma _{t}(A) &:&=Ad\Delta ^{it}(A)\in \mathcal{A} \\
&&AdJ(\mathcal{A})=\mathcal{A}^{\prime }  \notag
\end{eqnarray}
The modular automorphism group $\sigma _{t}$ depends only on the state $%
\omega (\cdot )=\left\langle \Omega \left| \cdot \right| \Omega
\right\rangle $ and not its implementing vector $\Omega ;$ they are related
through the KMS property (strip-analyticity of the function F(z)) 
\begin{eqnarray*}
F(t) &=&\omega (\sigma _{t}(A)B) \\
F(t+i) &=&\omega (B\sigma _{t}(A))
\end{eqnarray*}
The KMS property (which generalizes the Gibbs formula) characterizes the
modular automorphism of $\mathcal{A},\omega $
\end{theorem}

\begin{theorem}
(Bisognano-Wichmann) If we take for $\mathcal{A}$ the wedge algebra $%
\mathcal{A}=\mathcal{A}(W)$ and for $\Omega $ the vacuum vector, the modular
objects have the following physical interpretation 
\begin{eqnarray*}
\Delta ^{it} &=&U(\Lambda _{W}(-2\pi t)) \\
J &=&TCP\cdot Rot_{W}(\pi )
\end{eqnarray*}
\end{theorem}

Here $\Lambda _{W}(\chi )$ is the wedge adapted L-boost, $TCP$ the
antiunitary TCP-transformation of local QFT and $Rot_{W}(\varphi )$ the
rotation group around the spatial axis pointing into $W$.

\begin{definition}
(Wiesbrock, Borchers \cite{Bo}) An inclusion of operator algebras ($\mathcal{%
A}\subset \mathcal{B},\Omega $) is ''modular'' if ($\mathcal{A}$,$\Omega $),
($\mathcal{B}$,$\Omega $) are standard and $\Delta _{\mathcal{B}}^{it}$ acts
for (t\TEXTsymbol{<}0) as a compression on $\mathcal{A}$ 
\begin{equation}
Ad\Delta _{\mathcal{B}}^{it}\mathcal{A\subset A}
\end{equation}
A modular inclusion is standard if the relative commutant ($\mathcal{A}%
^{\prime }\cap \mathcal{B},\Omega $) is standard. In that case the
conditional expectation E
\end{definition}

\begin{remark}
In case the equality sign holds (i.e. the compression is an automorphism, a
theorem of Takesaki states the equality $\mathcal{B}$=$\mathcal{A}$.
Therefore the modular inclusion situation is a generalization of that
studied by Takesaki.
\end{remark}

\begin{theorem}
(Guido,Longo and Wiesbrock \cite{GLW}) Standard modular inclusions are in
correspondence with chiral AQFT
\end{theorem}

An important physical application used in the main text is obtained by the
following adaptation to a massive interacting AQFT 
\begin{eqnarray}
\mathcal{B} &=&\mathcal{A}(W) \\
\mathcal{A} &=&U(e_{+})\mathcal{A}(W)U^{\ast }(e_{+}),\,e_{+}=(1,1)  \notag
\\
&\equiv &\mathcal{A}(W_{e_{+}})  \notag
\end{eqnarray}
This is the inclusion of the algebra translated via a lightlike translation
into itself so the geometrically the relative commutator 
\begin{equation}
\mathcal{A}(W_{e_{+}})^{\prime }\cap \mathcal{A}(W)\equiv \mathcal{A}(I(0,1))
\end{equation}
is by causality localized in the upper horizontal interval (0,1). The
standardness of this inclusion then leads to a chiral conformal AQFT i.e. a
net (more precisely a pre-cosheaf) 
\begin{eqnarray}
I &\rightarrow &\mathcal{A}(I),\,\,I\subset S^{1} \\
\mathcal{A}(R_{+}) &=&\overline{\cup _{t}Ad\Delta ^{it}\mathcal{A}(I(0,1))} 
\notag \\
\mathcal{A}(R) &=&\mathcal{A}(R_{+})\vee J\mathcal{A}(R_{+})J
\end{eqnarray}
on which the Moebius group which preserves the vacuum vector acts. With the
help of an external (i.e. not in Moeb.) automorphism on $\mathcal{A}(R)$
implemented by the opposite lightray translation $U_{-}(a)$ we are able to
return from the chiral net on the right upper horizon to the original 2-dim.
net. We call this chiral theory supplemented by the opposite lightray
automorphism the holographic image of the 2-dim. massive net. With this
interpretation the holographic projection for massive d=1+1 theories is
nothing else than the conceptually and mathematically tightened version of
the old lightray/p$\rightarrow \infty $ frame quantization including the
rules of how to reprocess back the holographic image into the original local
d=1+1 theory. For the extension of holographic projection to higher
dimensional theory one needs one more mathematical definition and theorem
about ``modular intersections''

\begin{definition}
A ($\pm $) modular intersection is defined in terms of two standard pairs ($%
\mathcal{N}$,$\Omega $),$\,(\mathcal{M},\Omega )$ whose intersection is also
standard ($\mathcal{N\cap M},\Omega $) and which in addition fulfill 
\begin{eqnarray*}
&&\left( \left( \mathcal{N\cap M}\right) \subset \mathcal{N},\mathcal{\Omega 
}\right) \text{ }and\text{ }\left( \left( \mathcal{N\cap M}\right) \subset 
\mathcal{M},\mathcal{\Omega }\right) \,\,is\,\ \pm \,\func{mod}ular \\
&&J_{\mathcal{N}}(\lim_{t\rightarrow \mp \infty }\Delta _{\mathcal{N}%
}^{it}\Delta _{\mathcal{M}}^{-it})J_{\mathcal{N}}=\lim_{t\rightarrow \mp
\infty }\Delta _{\mathcal{M}}^{it}\Delta _{\mathcal{N}}^{-it}=J_{\mathcal{M}%
}\lim_{t\rightarrow \mp \infty }\Delta _{\mathcal{N}}^{it}\Delta _{\mathcal{M%
}}^{-it})J_{\mathcal{M}}
\end{eqnarray*}
\end{definition}

All limits are in the modular setting are to be understood in the sense of
strong convergence on Hilbert space vectors. In the geometric setting of
local quantum physics the the modular intersection property is realized par
excellence by the pair of intersecting wedge algebras $\mathcal{M=A(}W%
\mathcal{)},\mathcal{N}=AdU(\Lambda _{e_{+}}(a))\mathcal{M}$ together with
the $\Omega =$vacuum. Here $\Lambda _{e_{+}}(a)$ denotes a ``translation''
(transversal Galilei transformation) in the Wigner little group which fixes
the lightray vector $e_{+},$ i.e. the Lorentz transformation which tilts $%
\mathcal{W}$ around this lightray. In fact the limit in the second line is
geometrically nothing else but $\Lambda _{e_{+}}(a)$ and the commutation
relation with $J_{\mathcal{N},\mathcal{M}}$ is easily checked as a geometric
relation in the extended Lorentz group.

Modular intersections play an analogous role in the construction of 3- and
higher- dimensional AQFT starting from a finite set of wedge algebras \cite
{Kaehler} and the related holographic isomorphisms as the modular inclusions
used in section 3 for the 2-dimensional case.

\end{document}